\documentclass[12pt]{iopart}  
\usepackage{iopams}  
\usepackage{setstack}

\usepackage{amssymb}
\usepackage{braket} 
\usepackage{graphicx} 

\begin{document}

\title{Uniform Approximation from Symbol Calculus on a Spherical Phase Space}

\author{Liang Yu}

\address{ Department of Physics, University of California, Berkeley, California 94720 USA}
\ead{liangyu@wigner.berkeley.edu}

\begin{abstract}
We use symbol correspondence and quantum normal form theory to develop a more general method for finding uniform asymptotic approximations. We then apply this method to derive a result we announced in an earlier paper, namely, the uniform approximation of the $6j$-symbol in terms of the rotation matrices. The derivation is based on the Stratonovich-Weyl symbol correspondence between matrix operators and functions on a spherical phase space. The resulting approximation depends on a canonical, or area preserving, map between two pairs of intersecting level sets on the spherical phase space. 
\end{abstract}



\section{Introduction}

In this paper, we develop a more general method for finding uniform approximation using symbol calculus and quantum normal form theory. As an example, we use this more general method to derive a new uniform asymptotic formula for the Wigner $6j$-symbol in terms of Wigner rotation matrices. The asymptotic formula itself was announced without proof in our earlier paper \cite{littlejohn2009}. In this paper, we will fill in the proof, and leave out the implementation and numerical tests of this new uniform formula, which can be found in \cite{littlejohn2009}. 

The traditional method of finding uniform approximations is reviewed in Berry and Mount \cite{berry1972}. In the traditional approach for a Schr\"{o}dinger equation in one variable $x$, one uses a change of variable $X = X(x)$ to transform the original equation into a solvable one, called a ``comparison equation.'' One then expresses the asymptotic solution to the original equation in terms of the exact solution of the comparison equation. Because the action functions are the phase functions of the WKB formula, and because the WKB limit of the solution of the original equation and the WKB limit of the uniform formula must match, the requirement for the change of variable $X=X(x)$ is $s(x) = S(X)$, where $s(x)$ is the action for the original Hamiltonian, and $S(X)$ is the action for the comparison equation. These action functions $s(x)$ and $S(X)$ have geometrical interpretation in terms of phase space areas enclosed by the level sets of the classical Hamiltonian and the comparison equation, respectively. Geometrically, the condition $s(x)=S(X)$ requires the existence of a canonical, or area preserving map $X=X(x), P = (dx/dX) p$ that sends the level set of the original Hamiltonian to level set of the comparison equation. These special canonical maps, which are induced by coordinate transformations, are called point transformations. In other words, the traditional method of uniform approximation is based on the existence of point transformations. 

The more general approach to uniform approximations developed in this paper is based on the existence of more general canonical transformations. The main theoretical tool we use is a theory of quantum normal form, which transforms general differential operators into ``normal form'' operators based on the correspondence between canonical transformations and unitary transformations. This correspondence is developed using symbol calculus in \cite{cargo2005b, cargo2005a, cargo2002}. In this more general approach, we are no longer restricted to point transformations, or even differential equations. Since there exist more general symbol correspondence between matrix operators and functions on a sphere, we can apply the more general theory to matrix equations as well as differential equations. In this paper, we demonstrate this novel possibility by applying this method to derive a uniform approximation for the Wigner $6j$-symbol using the Stratonovich-Weyl symbol correspondence \cite{freidel2002, varilly1988, stratonovich1956} between finite dimensional matrix operators and functions on a spherical phase space. 

We now give an outline of this paper. In section \ref{ch5: sec_normal_form}, we review the basic result of quantum normal form theory, and recast the method of comparison equation in the language of a quantum normal forms. In section \ref{ch5: sec_6j_matrix_symbol}, we find the symbols for the matrix operators that define the Wigner $6j$-symbol, and look at the intersections of their associated level sets on the $2$-sphere. In section \ref{ch5: sec_determine_beta}, we choose the rotation matrices to be a normal form for the $6j$-symbol, and describe the canonical map used to construct the unitary transformation in section \ref{ch5: sec_symplectic_map}. Finally, in section \ref{ch5: sec_uniform_formula}, we derive the uniform approximation for the $6j$-symbol. The last section contains conclusions and comments.

\section{\label{ch5: sec_normal_form}Uniform Approximations From Quantum Normal Form Theory}

Quantum normal form theory is based on the $ \hbar $-expansion of a symbol correspondence, such as the Weyl symbol correspondence between operators on the Hilbert space ${\mathcal H} = L^2({\mathbb R})$ and functions on an ${\mathbb R}^2$ phase space. In such an $\hbar$-expansion, there is a natural relationship between operator commutators and Poisson brackets: 

\begin{equation}
\label{ch5: eq_commutator_bracket}
[ \hat{A} , \, \hat{B} ] \quad \leftrightarrow \quad     i \hbar \{A , \, B \}  + O ( \hbar^2) \, .
\end{equation}
In the above notation, a right arrow indicates the symbol of an operator, and a left arrow indicates the inverse symbol of a function. In general we will denote an operator with a hat on top, and its symbol by the same letter without the hat.

In equation (\ref{ch5: eq_commutator_bracket}), we can take $\hat{B}$ to be a Hamiltonian $\hat{H}$, and take $\hat{A}$ to be a generator for a unitary operator $\hat{U}$. Then integrating equation (\ref{ch5: eq_commutator_bracket}) gives us a relationship between a unitary transformation of the Hamiltonian $\hat{H}$ and a canonical transformation of the classical Hamiltonian $H$: 

\begin{equation}
\label{ch5: eq_normal_form_operator}
\hat{K} = \hat{U} \hat{H} \hat{U}^\dagger  \quad \leftrightarrow \quad K = H \circ Z^{-1} \, , 
\end{equation}
Here $\hat{K}$ is a new Hamiltonian, called a normal form of the original Hamiltonian $\hat{H}$, and $Z$ is a canonical transformation generated by the generator $A$.  The strategy is to choose $Z$ so that the normal form $\hat{K}$ is simple and solvable. This is the main idea behind quantum normal form theory. For more details on the construction of the unitary operator $\hat{U}$ from the canonical transformation $Z$, see \ref{ch5: section_unitary_operator}.

We now use the correspondence in equation (\ref{ch5: eq_normal_form_operator}) to find the wavefunction $\psi(x_0)$ of an eigenstate $\ket{\psi}$, where the eigenstate satisfies the eigenvalue equation
\begin{equation}
\label{ch5: eq_eigen_H}
(\hat{H} - E_0 ) \ket{\psi} = 0 \, , 
\end{equation}
and where $E_0$ is an eigenvalue of some arbitrary Hamiltonian $\hat{H}$.

\begin{figure}[tbhp]
\begin{center}
\includegraphics[width=0.90 \textwidth]{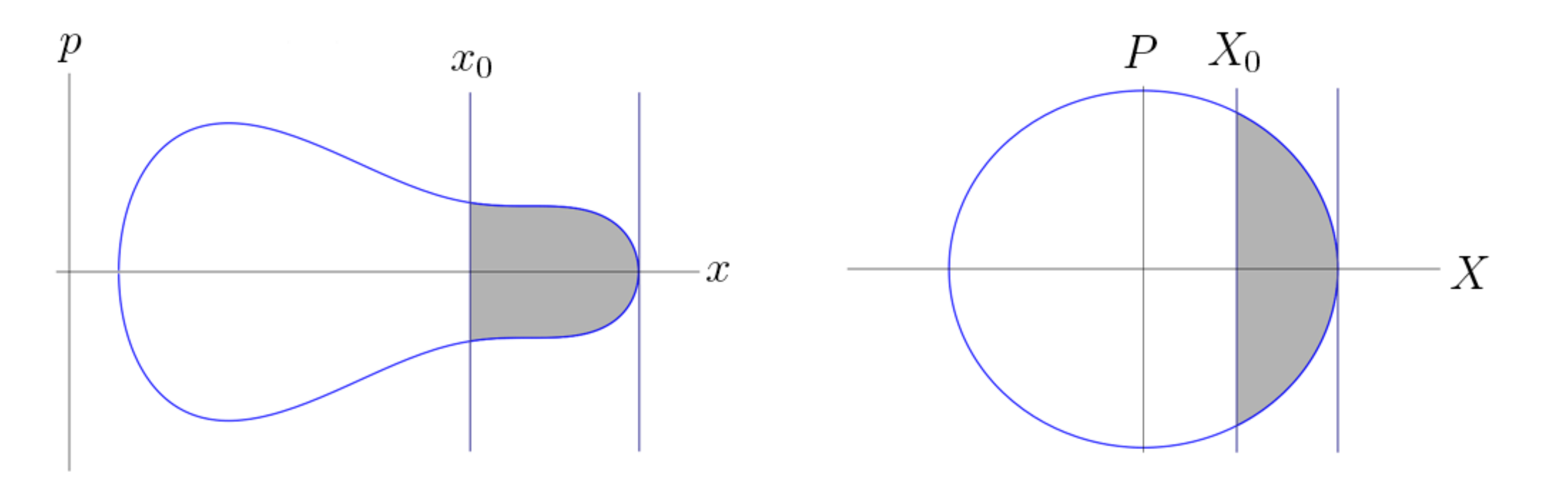}
\caption{The shaded areas in the left and right figures represent the action integrals $s(x_0)$ and $S(X_0)$, respectively, of the original Hamiltonian (on the left) and the normal form Hamiltonian (on the right). The normal form Hamiltonian is taken to be a quantum harmonic oscillator Hamiltonian.}
\label{ch5: fig_action_integral}
\end{center}
\end{figure}

The action integral $s(x_0) = \int_{x \ge x_0} p(x) \, dx$ is the area bounded by the level set $H(x, p) - E_0 = 0$ and the line $x = x_0$ in the $x$-$p$ phase plane. This area is illustrated on the left of figure \ref{ch5: fig_action_integral}. Suppose there is a canonical map $Z: (x, p) \rightarrow (X(x, p), P(x, p))$ that sends the vertical line $x = x_0$ to another vertical line $X = X_0$, and sends the level set $H - E_0 = 0$ to another level set $N(X, P) - N_0 = 0$, where $N(X, P)$ is the symbol of a normal form operator $\hat{N}$. Since this map is canonical, we must choose $X_0$ and $N$ so that the shaded area on the left of figure \ref{ch5: fig_action_integral}, which  represents the action $s(x_0)$, is equal to the shaded area on the right of figure \ref{ch5: fig_action_integral}, which represents the action $S(X_0) = \int_{X \ge X_0} P \, dX$. In other words, we suppose there exist a canonical map $Z$, a parameter $X_0$, and a function $N$ that satisfy the conditions

\begin{eqnarray}
\label{ch5: eq_Z_conditions_1}
H \circ Z^{-1} - E_0 &=& A ( N - N_0 )  \, , \\
\label{ch5: eq_Z_conditions_2}  
x \circ Z^{-1} - x_0 &=& B ( x - X_0 ) \, , 
\end{eqnarray}
for some functions $A = A(x, p)$ and $B= B(x, p)$. We construct a unitary operator $\hat{U}$ based on this canonical map $Z$, and define 

\begin{equation}
\hat{K}_1 = \hat{U} \hat{H} \hat{U}^\dagger \, ,  \quad \quad \quad 
\hat{K}_2 = \hat{U} \hat{x} \hat{U}^\dagger  \, .  
\end{equation}
From equation (\ref{ch5: eq_normal_form_operator}), the symbols of $\hat{K}_1$ and $\hat{K}_2$ are given by the right hand side of equation (\ref{ch5: eq_Z_conditions_1}) and equation (\ref{ch5: eq_Z_conditions_2}), respectively. We have

\begin{equation}
K_1 = A ( N - N_0 ) \, ,   \quad \quad \quad
K_2 = B ( x - X_0 )  \, .
\end{equation}
We now relate the eigenstate $\ket{\psi}$ of $\hat{H}$ to the eigenstate $\ket{N_0}$ of $\hat{N}$ through the unitary operator $\hat{U}$ and the non-unitary operator $\hat{A}$. First we insert $ \hat{U} $ on the left and the identity operator $ \hat{I} = \hat{U}^\dagger \hat{U}$ on the right of $(\hat{H} - E_0)$ in equation  (\ref{ch5: eq_eigen_H}), we find

\begin{equation}
\label{ch5: eq_K_1_transformation}
    \hat{U} (\hat{H} - E_0)  \hat{U}^\dagger \hat{U} \, \Ket{ \psi } \, = \hat{K}_1  \hat{U} \, \Ket{\psi } = 0 \, . 
\end{equation}
We then insert $ \sqrt{\hat{A}}$ on the left and the identity operator $ \hat{I} = \frac{1}{\sqrt{\hat{A}}} \, \sqrt{\hat{A}}
$ on the right of $(\hat{H} - E_0)$ in equation  (\ref{ch5: eq_K_1_transformation}), we find

\begin{equation}
\label{ch5: eq_A_U_psi}
    \sqrt{\hat{A}} \, (\hat{K}_1 ) \frac{1}{\sqrt{\hat{A}}} \sqrt{\hat{A}} \, \hat{U}  \, \Ket{ \psi } 
    \, = \hat{A} \, (\hat{N} - N_0 ) ( \sqrt{\hat{A}} \, \hat{U}  \, \Ket{ \psi } ) = 0 \, , 
\end{equation}
where we have used an operator identity 

\begin{equation}
 \sqrt{\hat{A}} \, (\hat{K}_1 ) \frac{1}{\sqrt{\hat{A}}} = \hat{A} \, (\hat{N} - N_0 ) \, . 
 \label{ch5: eq_op_identity}
\end{equation}
This operator identity equation (\ref{ch5: eq_op_identity}) follows from a general star product identity that holds to first order in $\hbar$, 

\begin{equation}
\label{ch5: eq_star_prod_identity}
    \sqrt{A} \, * ( A (N - N_0) ) * \frac{1}{\sqrt{A}} = A * ( N - N_0 )   \, .
\end{equation}
From equation (\ref{ch5: eq_A_U_psi}), we conclude that 

\begin{equation}
\label{ch5: eq_N_psi}
\ket{N_0} =  \sqrt{\hat{A}} \, \hat{U} \, \ket{\psi}  \, ,
\end{equation}
Inverting the relationship between $\ket{\psi}$ and $\ket{N_0}$ in equation  (\ref{ch5: eq_N_psi}), we find

\begin{equation}
\label{ch5: eq_normal_form_psi_ket}
\ket{\psi} = \hat{U}^\dagger \frac{1}{\sqrt{\hat{A}}} \ket{N_0} \, . 
\end{equation}
An analogous calculation for the operator $\hat{x}$ leads to  

\begin{equation}
\label{ch5: eq_normal_form_x_ket}
  \ket{x_0} = \hat{U}^\dagger \frac{1}{\sqrt{\hat{B}}} \Ket{X_0}  \, , 
\end{equation}
where $\Ket{X_0}$ is an eigenstate of $\hat{x}$ with eigenvalue $X_0$.

Taking the scalar product between the states in equation (\ref{ch5: eq_normal_form_psi_ket}) and equation (\ref{ch5: eq_normal_form_x_ket}), we get the wavefunction of the eigenstate $\psi(x_0) = \braket{x_0 | \psi}$, 
\begin{eqnarray}
  \Braket{x_0 | \psi } &=& \Braket{X_0 |\, \frac{1}{\sqrt{\hat{B}}} \,\hat{U} \, \hat{U}^\dagger \frac{1}{\sqrt{\hat{A}}} \, | N_0 } \, ,    \nonumber   \\     
  &=&   \Braket{X_0 |\, \frac{1}{\sqrt{\hat{B}}} \, \frac{1}{\sqrt{\hat{A}}} \, | N_0 } \, , 
    \label{ch5: eq_general_uniform}
\end{eqnarray}
where we have used the unitary property to cancel $\hat{U}$ and $\hat{U}^\dagger$ in the second equality.

As an example, let us use equation (\ref{ch5: eq_general_uniform}) to derive the Airy type uniform approximation for a general Schr\'{o}dinger equation, which has the following form

\begin{equation}
( \hat{p}^2 - p^2(\hat{x}) ) \ket{\psi} = 0 \, , 
\end{equation}
where $p(\hat{x}) = \sqrt{E_0 - V(\hat{x})}$, and $V(x)$ is some general potential function. We will assume that there exist a canonical map $Z$, given by ($X = F(x)$, $P = p / F'(x)$), that sends the level set of the symbol of the Schr\"{o}dinger equation, $p^2 - p^2(x) = 0$, into the level set of the symbol of the Airy equation, $P^2 + X = 0$. That is, we assume

\begin{equation}
(p^2 - p^2(x)) \circ Z^{-1} = (F'(x))^2 \left ( P^2 - \frac{p^2(x)}{(F'(x))^2} \right) =  A(X) (P^2 - X) \, .
\label{ch5: eq_airy_symbol_map}
\end{equation}
equation (\ref{ch5: eq_airy_symbol_map}) implies

\begin{equation}
A(X)  = (F'(F^{-1}(X)))^2 \, , 
\label{ch5: eq_A_expression}
\end{equation}

\begin{equation}
F'(F^{-1}(X)) = p(F^{-1}(X)) / \sqrt{X} \, , 
\label{ch5: eq_F_prime_expression}
\end{equation}
From the relation $dX/dx = F'(x) = p(x) / \sqrt{X}$, we deduce   

\begin{equation}
 \int p \,  dx  = \frac{2}{3} X^{3/2} \, .
\end{equation}
This is the standard change of variable used in the Airy type uniform approximation.

Since we are using a point transformation, the explicit form of the unitary transformation is available. It is simply given by $\hat{U} \ket{x_0} = \sqrt{F'(x_0)} \ket{X_0}$. Thus we have

\begin{equation}
\ket{x_0} = \hat{U}^\dagger \, ( \sqrt{F'(F^{-1}(\hat{x}))} ) \ket{X_0} \, , 
\label{ch5: eq_x_0_state}
\end{equation}
By comparing equation (\ref{ch5: eq_x_0_state}) with equation (\ref{ch5: eq_normal_form_x_ket}), we have

\begin{equation}
\hat{B} = \frac{1}{F'(F^{-1}(\hat{x}))} \, , 
\label{ch5: eq_B_expression}
\end{equation}
Finally, using equation (\ref{ch5: eq_general_uniform}), we obtain 

\begin{eqnarray}
\psi(x_0) &=& \Braket{X_0 |\, \frac{1}{\sqrt{\hat{B}}} \, \frac{1}{\sqrt{\hat{A}}} \, | N_0 }  \nonumber  \\ 
&=& \Braket{X_0 |\, \frac{1}{\sqrt{F'(F^{-1}(\hat{x}))}}  \, | N_0 }  \nonumber   \\ 
&=& \Braket{X_0 |\, \frac{1}{\sqrt{p(F^{-1}(\hat{x})) / \sqrt{ \hat{x} } }}  \, | N_0 }   \nonumber  \\ 
&=& \frac{X_0^{1/4}}{ (p(x_0))^{1/2} } \, {\rm Ai} (X_0)  \, ,
\end{eqnarray}
where we have used equation (\ref{ch5: eq_A_expression}) and equation (\ref{ch5: eq_B_expression}) in the second equality, and equation (\ref{ch5: eq_F_prime_expression}) in the third equality. This is the result for the Airy type uniform approximation for a general Schr\"{o}dinger equation, so the more general method developed here is consistent with the result from the traditional method of comparison equations. We now turn our attention to the $6j$-symbol, where a different symbol correspondence is required, and where point transformations are not possible on a sphere.

\section{\label{ch5: sec_6j_matrix_symbol}Matrix Operators and Symbols on a Spherical Phase Space}

In the rest of this paper, we use equation (\ref{ch5: eq_general_uniform}) to derive a uniform approximation for the $6j$-symbol in terms of the $d$-matrices. The result itself was announced in equation (69) from our earlier paper \cite{littlejohn2009}. It has the form

\begin{eqnarray}
    \left\{ 
   \begin{array}{ccc} 
     j_1 & j_2 & j_{12} \\
     j_3 & j_4 & j_{23} 
   \end{array}
   \right\} 
   &=& (-1)^{\gamma} \, \left(\frac{ | \sin \beta \, J_y | }{ | 24V | } \right)^{1/2} \; d^j_{mm'} (\beta)
   \label{ch5: eq_uniform_formula_1}
\end{eqnarray}
where $\beta$ is implicitly defined in equation (\ref{ch5: eq_beta_eqn}), $J_y$ and $V$ are defined below equations (\ref{ch5: eq_Jz_Jn_bracket}) and (\ref{ch5: eq_J12_J23_bracket}), respectively, and $\gamma$ is defined in equations (69) and (75) of \cite{littlejohn2009}. 

Our strategy is to convert matrix operators that define the $6j$-symbol into functions on a spherical phase space using the Stratonovich-Weyl symbol correspondence, then show that the operators which define the $d$-matrices are suitable normal forms, and finally apply equation (\ref{ch5: eq_general_uniform}) to derive the uniform approximation. 

We set $\hbar = 1$, so all angular momenta are dimensionless. Consider the Hilbert space generated from the tensor product of four angular momentum spaces, associated with the operators $\hat{{\bf J}}_1, \hat{{\bf J}}_2, \hat{{\bf J}}_3, \hat{{\bf J}}_4$, and the corresponding eigenvalues $j_1, j_2, j_3, j_4$, respectively. Assuming the eigenvalues satisfy the triangle inequalities, there is a one dimensional subspace $Z$ that is the zero eigenspace of the total angular momentum operator, 

\begin{equation}
\label{ch5: eq_total_J_0}
 \hat{{\bf J}}_1 + \hat{{\bf J}}_2 + \hat{{\bf J}}_3 + \hat{{\bf J}}_4 = {\bf 0} \, .
\end{equation}
In this subspace $Z$, there are two basis sets labeled by the intermediate angular momenta $j_{12}$ and $j_{23}$, respectively. The quantum numbers $j_{12}$ and $j_{23}$ are respectively the eigenvalues of the squares of the operators

\begin{equation}
\hat{\bf J}_{12} = \hat{\bf J}_1 + \hat{\bf J}_2 \, ,    \quad \quad \quad 
\hat{\bf J}_{23} = \hat{\bf J}_2 + \hat{\bf J}_3  \, . 
\end{equation}
The $6j$-symbol is proportional to the unitary matrix element $\braket{j_{12} | j_{23}}$ that represents the change of basis from $\ket{j_{12} }$ to $\ket{ j_{23} }$ in $Z$. The definition of the $6j$-symbol \cite{edmonds1960} is given by 

\begin{equation}
\label{ch5: eq_6j_symbol_def}
    \left\{ 
   \begin{array}{ccc} 
     j_1 & j_2 & j_{12} \\
     j_3 & j_4 & j_{23} 
   \end{array}
   \right\}  = \frac{1}{\sqrt{(2j_{12} + 1)(2j_{23}+1)}} \braket{j_{12} | j_{23}}  \, . 
\end{equation}
To be defined, the $6j$-symbol must satisfy four triangle inequalities, in $(j_1, j_2, j_{12})$, $(j_2, j_3, j_{23})$, $(j_3, j_4, j_{12})$, and $(j_1, j_4, j_{23})$. For example, $j_{12}$ must lie between the bounds

\begin{equation}
| j_1 - j_2 | \le j_{12} \le j_1 + j_2
\end{equation}
in integer steps. For given $j_i$, $i = 1,2,3,4$, these imply that $j_{12}$ and $j_{23}$ vary between the limits 

\begin{equation}
\label{ch5: eq_j_12_inequality}
j_{12, {\rm min}}  \le j_{12} \le j_{12, {\rm max}} \, ,  \\
j_{23, {\rm min}}  \le j_{23} \le j_{23, {\rm max}}  \, , 
\end{equation}
in integer steps, where 

\begin{eqnarray}
j_{12, {\rm min}} = \max ( |j_1 - j_2| , |j_3 - j_4| )  \, ,  \\
j_{23, {\rm min}} = \max ( |j_2 - j_3| , |j_1 - j_4| )  \, , \\
j_{12, {\rm max}} = \min ( j_1 + j_2, j_3 + j_4 )  \, , \\
j_{23, {\rm max}} = \min ( j_2 + j_3, j_1 + j_4  )  \, .
\end{eqnarray}
We shall reserve lower case $j_i$ for the quantum numbers, and for the purposes of symbol calculus, we shall set 

\begin{equation}
\label{ch5: eq_J_j_relation}
J_i = j_i + \frac{1}{2} \, ,
\end{equation}
for $i = 1,2,3,4,12,23$. It is useful to rewrite equation (\ref{ch5: eq_j_12_inequality}) for these variables. We find $J_{12}$ and $J_{23}$ vary between the limits 

\begin{equation}
\label{ch5: eq_classical_J_12_inequality}
J_{12, {\rm min}}  \le J_{12} \le J_{12, {\rm max}} \, ,  \\
J_{23, {\rm min}}  \le J_{23} \le J_{23, {\rm max}}  \, , 
\end{equation}
in integer steps. Here 

\begin{eqnarray}
J_{12, {\rm min}} = \max ( |J_1 - J_2| , |J_3 - J_4| ) \, ,  \\
J_{23, {\rm min}} = \max ( |J_2 - J_3| , |J_1 - J_4| ) \, ,  \\
J_{12, {\rm max}} = \min ( J_1 + J_2, J_3 + J_4 )  \, ,  \\
J_{23, {\rm max}} = \min ( J_2 + J_3, J_1 + J_4  ) \, .
\end{eqnarray}
The number of allowed $j_{12}$ or $j_{23}$ values is the same, and it is the dimension $D$ of the subspace $Z$ as well as the size of the matrix $\braket{j_{12} | j_{23} }$.

\begin{equation}
D  =  \dim Z = j_{12, {\rm max}} - j_{12, {\rm min}} + 1 = j_{23, {\rm max}} - j_{23, {\rm min}} + 1 \, . 
\label{ch5: eq_D_dim}
\end{equation}
The dimension of $Z$ can be written as $D = 2j + 1$ for some integer or half integer $j$. The Stratonovich-Weyl symbol correspondence provides a mapping between $2j+1$ dimensional matrices and functions on a $2$-sphere of radius $J = j+1/2$. This mapping depends on the basis we choose in $Z$. In order to fix the symbol correspondence, we shall choose the $\ket{j_{12}}$ basis, which we relabel to a standard basis used in the symbol correspondence,  

\begin{equation}
\ket{j, m} = \ket{j_{12}}  \, , 
\end{equation}
where 

\begin{equation}
j = (D - 1) /2 \, , 
\end{equation}
and

\begin{equation}
\label{ch5: eq_m_def}
m = j_{12} - 1/2 ( j_{12, {\rm min}} + j_{12, {\rm min}} )  = - j, \dots, j 
\end{equation}
is the deviation of $j_{12}$ from its average value. Using the $\ket{j, m}$ basis for the symbol correspondence, we define the generators $\hat{K}_x, \hat{K}_y, \hat{K}_z$ that satisfy the commutation relations of the $SU(2)$ algebra as follows. Let $\hat{K}_+$ and $\hat{K}_-$ be the usual raising the lower operators for the  $\ket{j, m }$ basis, respectively. Define 

\begin{eqnarray}
\hat{K}_x = \frac{1}{2} \, ( \hat{K}_+  +  \hat{K}_- ) \, ,  \\
\hat{K}_y =  \frac{1}{2 i } \, ( \hat{K}_+  -  \hat{K}_- ) \, ,  \\
\hat{K}_z = \hat{J}_{12} - J_{12, {\rm avg}}  \, ,
\label{ch5: eq_K_z_operator}
\end{eqnarray}
where 

\begin{equation}
\hat{J}_{12} = \sqrt{ \hat{\bf J}_{12}^2 } \, , 
\end{equation}

\begin{equation}
J_{12, {\rm avg}}  = \frac{1}{2} \, ( J_{12, {\rm min}} + J_{12, {\rm max}} ) \, .
\end{equation}
The Stratonovich Weyl symbol of the operators $(\hat{K}_x, \hat{K}_y, \hat{K}_z)$ are given by the Cartesian coordinate functions $(K_x, K_y, K_z)$ of a $2$-sphere of radius $J = D/2 = j+1/2$, with spherical angles $(\theta_{12}, \phi_{12})$. Explicitly, 

\begin{eqnarray}
K_x = (D/2) \sin \theta_{12} \, \cos \phi_{12}  \, ,  \\
K_y = (D/2) \sin \theta_{12} \, \sin \phi_{12}  \, ,  \\
K_z = (D/2) \cos \theta_{12} \, .
\label{ch5: eq_K_z_symbol}
\end{eqnarray}
We have labeled the azimuthal angle by $\phi_{12}$, since it is the conjugate angle to $J_{12}$. 

\begin{figure}[tbhp]
\begin{center}
\includegraphics[width=0.50 \textwidth]{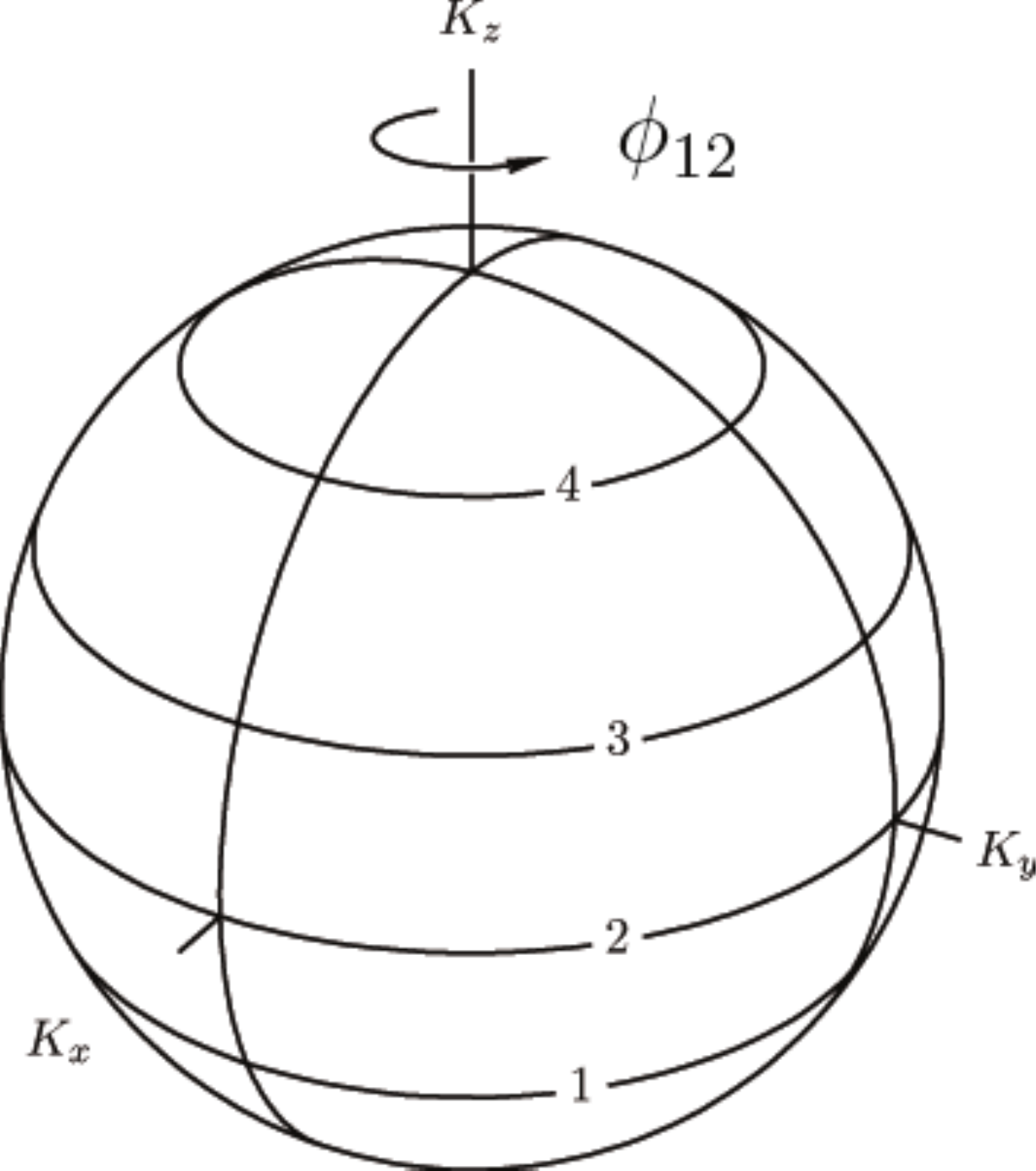}
\caption{The phase space of the $6j$-symbol is a sphere of radius $D/2$ in a space in which $(K_x, K_y, K_z)$ are Cartesian coordinates. To within an additive constant, $K_z$ is $J_{12}$. Several curves of constant $J_{12}$, which are small circles, are shown}
\label{ch5: fig_K_z_level_set}
\end{center}
\end{figure}

We call this spherical phase space the $6j$-sphere. On the $6j$-sphere, the north pole is at $K_z = J$, or $J_{12} = J_{12, {\rm max}}$, and the south pole is at $K_z = - J$, or $J_{12} = J_{12, {\rm min}}$, and curves of constant $J_{12}$ in general are small circles $K_z = {\rm const}$. The $6j$-sphere is illustrated in figure \ref{ch5: fig_K_z_level_set}, which shows several curves of constant $J_{12}$. Our choice of the basis $\ket{j_{12}}$ for the symbol correspondence makes the level sets of $J_{12}$ very nice. The level sets of $J_{23}$, however, will not in general be small circles.

The states $\ket{j_{12}}$ and $\ket{j_{23}}$ are defined by the operator equations

\begin{eqnarray}
\left( \hat{J}_{12} - \sqrt{ j_{12} ( j_{12} + 1)} \right)  \ket{j_{12}} = 0 \, ,  \\
\left( \hat{J}_{23} - \sqrt{ j_{23} ( j_{23} + 1)} \right)  \ket{j_{23}} = 0 \, , 
\end{eqnarray}
where $\hat{J}_{23} = \sqrt{ \hat{\bf J}_{23}^2 }$. In the asymptotic limit of large $j$'s, we can replace $\sqrt{j_{12} (j_{12} + 1)}$ and $\sqrt{j_{23} (j_{23} + 1)}$ by $J_{12} = j_{12} + 1/2$ and $J_{23} = j_{23} + 1/2$, respectively. The operator equations defining $\ket{j_{12}}$ and $\ket{j_{23}}$ become

\begin{eqnarray}
\left( \hat{J}_{12} - (j_{12} + 1/2) \right)  \ket{j_{12}} = \left( \hat{K}_z - m \right) \ket{j_{12}} = 0 \, ,  \\
\left( \hat{J}_{23} - ( j_{23} + 1/2) \right)  \ket{j_{23}} = 0 \, , 
\end{eqnarray}
where we have used the definitions of $\hat{K}_z$ from equation (\ref{ch5: eq_m_def}) and the definition of $m$ from equation (\ref{ch5: eq_K_z_operator}) in the first equation. We now find the symbols of the operators $\hat{K}_z$ and $\hat{J}_{23}$. The symbol $K_z$ of $\hat{K}_z$ is given by equation (\ref{ch5: eq_K_z_symbol}) above. The exact matrix elements of the operator $\hat{J}_{23}^2$ in the $\ket{j_{12}} = \ket{j, m}$ basis can be found from the appendix of \cite{schulten1975a}. It is a symmetric tridiagonal matrix, so it has the form 

\begin{equation}
	\hat{J_{23}^2} =  \hat{A} +  \frac{1}{2} (\hat{J}_- \hat{B} + \hat{B} \hat{J}_+ )  \, ,
\end{equation}
where $\hat{A}$ and $\hat{B}$ can be read off from the following matrix elements:

\begin{eqnarray}
 \Braket{j_{12} | \hat{J}_{23}^2 | j_{12}}
	&& = \frac{1}{2j_{12}(j_{12}+1)} \{ j_{12}(j_{12}+1)[-j_{12}(j_{12}+1) + j_1(j_1+1)+j_2(j_2+1)]   \nonumber \\
	 && \quad \quad + j_3(j_3+1)[j_{12}(j_{12}+1) + j_1(j_1+1)-j_2(j_2+1)] \nonumber \\
	 &&  \quad \quad  + j_4(j_4+1)[j_{12}(j_{12}+1) - j_1(j_1+1)+j_2(j_2+1)] \}  \, , 
\end{eqnarray}

\begin{eqnarray}
  &&\Braket{ j_{12}-1 | \hat{J}_{23}^2 | j_{12}} \\
  &=& \frac{\left\{[j_{12}^2 - (j1-j2)^2][(j1+j2+1)^2-j_{12}^2][j_{12}^2 - (j3-j4)^2][(j3+j4+1)^2-j_{12}^2]\right\}^{1/2}}{2j_{12}[(2j_{12}-1)(2j_{12}+1)]^{1/2}}  \, .  \nonumber
\end{eqnarray}
Note that $\hat{A}$ and $\hat{B}$ only depend on $\hat{J}_{12}$, or $\hat{K}_z$.

Making the approximation $(2j_{12}+1)(2j_{12}-1) = 4 (j_{12}+1)^2$ in
the denominator of the off-diagonal matrix element, and noting that the
symbol of $\hat{K}_x = (\hat{J}_+ + \hat{J}_-)/2$ is $K_x$, we have
the symbol of $\hat{J}_{23}$ to first order

\begin{eqnarray}
	J_{23}^2 (J_{12}, \phi_{12})
	&=& \frac{1}{2 J_{12}^2} \left[ (J_{12}^2(-J_{12}^2+J_1^2+J_2^2) + J_3^2 (J_{12}^2+J_1^2-J_2^2) + J_4^2(J_{12}^2-J_1^2+J_2^2)\right]  \nonumber  \\
	&+& \frac{8}{J_{12}^2} \, F(J_{12}, J_1, J_2) \, F(J_{12}, J_3, J_4) \, \cos \phi_{12}  \, , 
	\label{ch5: eq_J_23_function}
\end{eqnarray}
where 

\begin{equation}
  F(a,b,c)= \frac{1}{4} [(a+b+c)(-a+b+c)(a-b+c)(a+b-c)]^{1/2}  
\end{equation}
is the area of a triangle with sides $a$, $b$, $c$.

\begin{figure}[tbhp]
\begin{center}
\includegraphics[width=0.40 \textwidth]{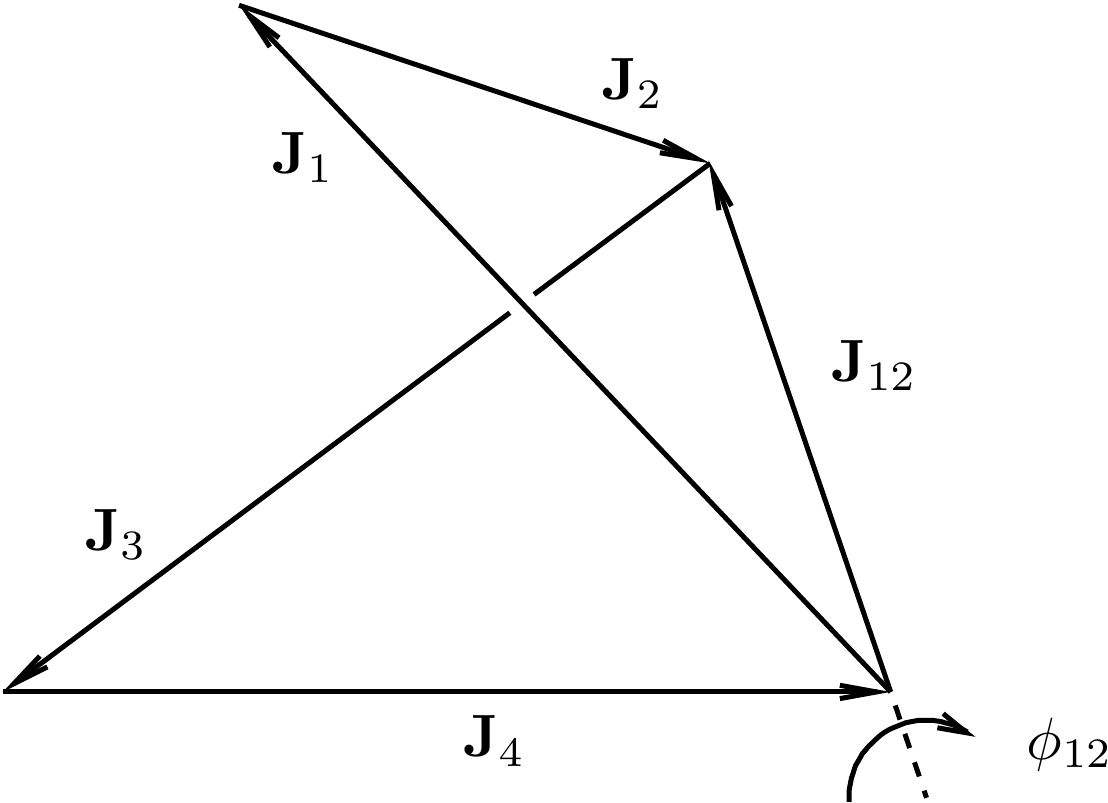}
\caption{The function $J_{23}$ is the sixth edge length, not shown, as a function of the five edge lengths shown and the dihedral angle $\phi_{12}$. }
\label{ch5: fig_tetrahedron_phi_12}
\end{center}
\end{figure}

It turns out that the function $J_{23}$ in equation  (\ref{ch5: eq_J_23_function}) is equal to the edge length $J_{23}$ as a function of the five edge lengths $J_1, J_2, J_3, J_4, J_{12}$ and the dihedral angle $\phi_{12}$ at the edge $J_{12}$ in a tetrahedron. See figure \ref{ch5: fig_tetrahedron_phi_12}, where the geometry of a tetrahedron formed with the six edges $J_1, J_2, J_3, J_4, J_{12}, J_{23}$ and dihedral angle $\phi_{12}$ is illustrated.

\pagebreak

\section{\label{ch5: sec_determine_beta}A Quantum Normal Form on a Spherical Phase Space}

\begin{figure}[tbhp]
\begin{center}
\includegraphics[width=0.95 \textwidth]{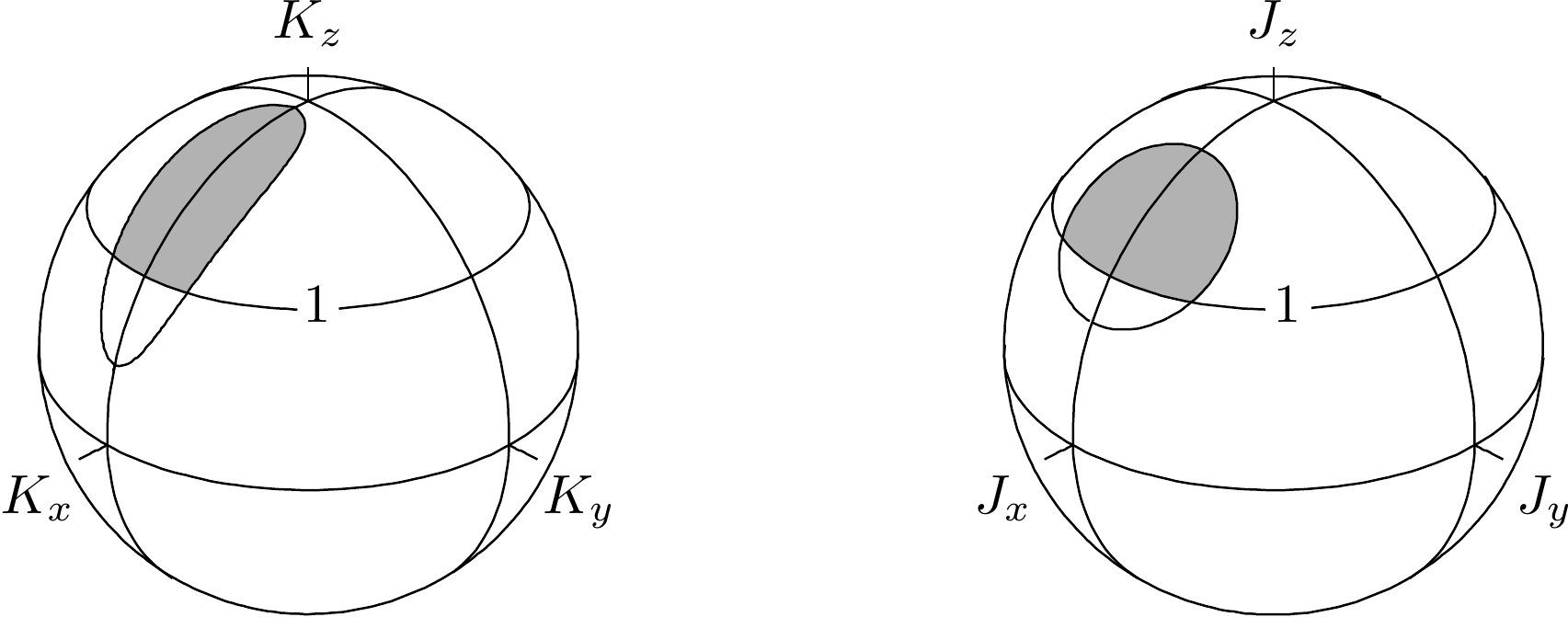}
\caption{The canonical map that maps the level set $J_{23} = j_{23}$ to a small circle $J_\beta = {\rm const}$ that is tilted by some angle $\beta$ around the $y$-axis. }
\label{ch5: fig_canonical_map_1}
\end{center}
\end{figure}

The operator $\hat{J}_{23}$ is analogous to the operator $\hat{H}$ in section \ref{ch5: sec_normal_form}. In order to apply the method from section \ref{ch5: sec_normal_form} to find a uniform approximation for the $6j$-symbol, we need to find a normal form operator $\hat{N}$, whose symbol $N$ has level sets that are similar to those of $J_{23}$. A good choice is $\hat{N} = \hat{J}_n$, where

\begin{equation}
\hat{J}_n = \hat{\bf n} \cdot {\bf \hat{J}} = \hat{J}_z \cos \beta + \hat{J}_y \sin \beta \, ,
\label{ch5: eq_J_n_op_def}
\end{equation}
and where $\hat{\bf n}$ is a unit vector in the $z$-axis that is rotated by an angle $\beta$ about the $y$-axis. The operators $\hat{J}_x, \hat{J}_y, \hat{J}_z$, like the operators $\hat{K}_x, \hat{K}_y, \hat{K}_z$, satisfy the $SU(2)$ generators commutation relations. Their symbols are the coordinate functions on another spherical phase space, which we call the $d$-sphere. The topology of the intersections of the $J_{12}$ and $J_{23}$ level sets on the $6j$-sphere are similar to the topology of the circle intersections of the $J_z$ and $J_n$ level sets on the $d$-sphere. This is illustrated in figure \ref{ch5: fig_canonical_map_1} and is described in detail in  \cite{littlejohn2009}.  Let $\ket{m'}$ denote the eigenstate of $\hat{J}_n$, then the eigenfunction of $\hat{J}_n$ is expressed in terms of the rotation matrices $d^j_{mm'}(\beta) = \braket{m|m'}$. The strategy is to express the $6j$-symbol, which is proportional to $\braket{j_{12}|j_{23}} = \braket{m|j_{23}}$, in terms of the rotation matrices $d^j_{mm'}(\beta) = \braket{m|m'}$. 

As described in detail in section 4.2 of \cite{littlejohn2009}, we can use the canonical condition to fix the parameters $j, m, m'$, and $\beta$ for the normal form. First, the area of the two spheres must be equal, which is equivalent to the condition that the dimensions of the matrix operators must be equal. Thus, $j$ is determined by 

\begin{equation}
j = (D-1)/2 \, , 
\end{equation}
where $D$ is defined in equation (\ref{ch5: eq_D_dim}). The level set of $J_{12} = j_{12} + 1/2$ is mapped to the level set $J_z = m$. Both level sets are small circles and they must contain the same area about the north pole. Thus we must have $J_{12, \, {\rm max}} - j_{12} = j - m$, or 

\begin{equation}
m = j_{12} - j_{12, \, {\rm avg}} \, , 
\end{equation}
where $j_{12, \, {\rm avg}} = (j_{12, \, {\rm min}} + j_{12, \, {\rm max}}) / 2$. Similarly, the area enclosed by the $J_{23}$ level set must equal to area enclosed by the small circle $J_{\hat{\bf n}} = m'$. This requires $j_{23} - j_{23, \, {\rm min}} = j - m'$, or 

\begin{equation}
m' =  j_{23, \, {\rm avg}} - j_{23} \, .
\end{equation}
Finally, the parameter $\beta$ is determined by equating the areas of the lunes indicated by the shaded regions in figure  \ref{ch5: fig_canonical_map_1}. The area of the lune on the $6j$-sphere is the integral $2 \int \phi_{12} \, dJ_{12}$, where $\phi_{12}$ is the coordinate on the $J_{23}$ level set. The identification of the function $J_{23}$ with the edge length in a tetrahedron allows us to use the Schl\"{a}fli identity \cite{milnor1994} to evaluate this integral. The Schl\"{a}fli identity states

\begin{equation}
\frac{d}{d J_{12}} \,  (\sum_{i=1}^6 J_i \psi_i )  = \psi_{12} = \pi - \phi_{12} \, , 
\end{equation}
where $\psi_i$ are exterior dihedral angles of the tetrahedron. Rearranging, we have

\begin{equation}
 \phi_{12} = \pi -  \frac{d}{d J_{12}} \,  (\sum_{i=1}^6 J_i \psi_i )  \, . 
\end{equation}
Integrating, we find

\begin{equation}
2 \int_{J_{12}}^{J_{12\, , {\rm max}}}  \phi_{12} \, dJ_{12} = 2 (J_{12, \, {\rm max}} - J_{12}) \pi  + 2 \Phi_{PR} - {\rm const} \, ,  
\end{equation}
where $\Phi_{\rm PR}= \sum_{i=1}^6 J_i \psi_i $ is the Ponzano-Regge phase evaluated at the intersection of the $J_{12}$ level set and the $J_{23}$ level set. The constant is derived in equation (74) of \cite{littlejohn2009}. The area of the shaded region in the sphere on the right of figure  \ref{ch5: fig_canonical_map_1} is given by the similar integral $2 \int \phi \, dJ_z$. This area is derived in equation (54) of \cite{littlejohn2009}. The result is denoted by $2 \Phi_d(\beta)$ and is a function of $\beta$. Equating the two areas, we obtain a condition for the parameter $\beta$, 

\begin{equation}
\Phi_{\rm PR} - \Phi_0 = \Phi_d (\beta)  \, , 
\label{ch5: eq_beta_eqn}
\end{equation}
where the constant $\Phi_0$ is derived in \cite{littlejohn2009} and is given by 

\begin{equation}
\Phi_0 = (J_1 + J_2 + J_3 + J_4 + J_{12} - J_{12, \, {\rm max}}) \, \pi \, . 
\end{equation}
By analytic continuation on both sides of equation (\ref{ch5: eq_beta_eqn}), the parameter $\beta$ can be determined in the classically forbidden region as well, as shown in equation (78) in  \cite{littlejohn2009}. For more details on the determination of $\beta$, see \cite{littlejohn2009}.

\section{\label{ch5: sec_symplectic_map}The Canonical Map on the Sphere}

\begin{figure}[tbhp]
\begin{center}
\includegraphics[width=0.95 \textwidth]{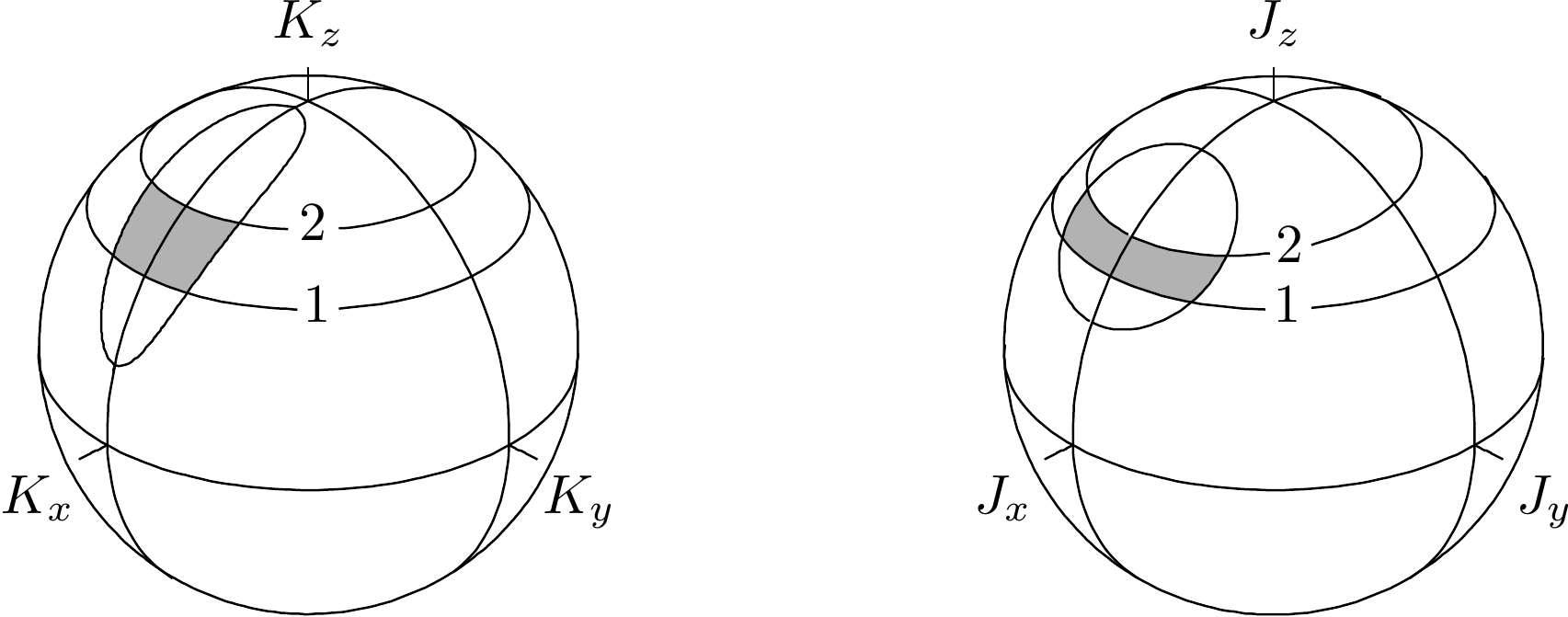}
\caption{The canonical map that maps a small circle $K_z = z$ to a small circle $J_\alpha = {\rm const}$ that is tilted by some angle $\alpha(z)$ around the $y$-axis. This tilt of the circle 2 effectively squeezes the shaded area in the $6j$-sphere, and pushes the curve $J_{23} = {\rm const}$ outward toward the curve $J_n = {\rm const}$ in the $d$-sphere.}
\label{ch5: fig_canonical_map}
\end{center}
\end{figure}

After we have fixed the value of $\beta$, we construct the required canonical map by sending the other small circles of $K_z = {\rm const}$ on the $6j$-sphere to tilted small circles on the $d$-sphere. The theory of uniform approximation requires that the canonical map sends the two curves $J_{12} = {\rm const}$ and $J_{23} = {\rm const}$ on the $6j$-sphere to the two curves $J_z = {\rm const}$ and $J_n = {\rm const}$ on the $d$-sphere, as illustrated in figure \ref{ch5: fig_canonical_map_1}. If the phase space is a plane instead of a sphere, and if the level sets of $J_{12} = {\rm const}$ are straight lines instead of small circles, we can squeeze or stretch the curve $J_{23} = {\rm const}$ along the $\phi_{12}$ direction into any desired shape. This can be done by moving the small circles, such as circle 2 on the left of figure \ref{ch5: fig_canonical_map}, up and down vertically. A point transformation uses this exact mechanism to deform the level set of an arbitrary Hamiltonian into a normal form. Since the phase space is a sphere of finite area, however, it is impossible to preserve the area between circle 1 and circle 2 while moving circle 2 up and down on the left of figure \ref{ch5: fig_canonical_map}. This is why it was necessary to generalize the theory of uniform approximation to handle more general canonical transformations that are not point transformations in the beginning.

Since we are allowed to use general canonical maps in our new approach, we can effectively squeeze or stretch the  shaded area in figure \ref{ch5: fig_canonical_map} by tilting circle 2, $K_z = c$, by some angle $\alpha(c)$ about the $y$ axis, as illustrated on the right of figure \ref{ch5: fig_canonical_map}. The angle $\alpha(c)$, which is a function of $c$, is determined by equating the area of the two shaded regions in figure \ref{ch5: fig_canonical_map}. The shaded area on the left is twice the difference of two Ponzano Regge actions $\Phi_{PR}$ for two different values of $J_{12}$. The area on the right is twice the difference between two values of $\Phi_d$ for different values of $m$ and $\beta$. For circle 1, $K_z = c_1$ or $J_{12} = j_{12} + 1/2$, the angle $\beta$ is given by equation (\ref{ch5: eq_beta_eqn}). For circle 2, $K_z = c$, the angle is decreased by $\alpha(c)$ from the tilt of the circle 2, so the angle is $\beta - \alpha(c)$. Thus, the equation for $\alpha(c)$ is given by 

\begin{equation}
\Phi_{\rm PR} (c) - \Phi_{\rm PR} (c_1) 
= \Phi_{\rm d}(\beta - \alpha(c), c) - \Phi_{\rm d}(\beta, c_1) \,  .
\end{equation}
By solving for $\alpha(c)$, we can construct a canonical map that sends the curves $J_{12} = {\rm const}$ and $J_{23} = {\rm const}$ to the curves $J_z = {\rm const}$ and $J_n = {\rm const}$.

\section{\label{ch5: sec_uniform_formula}A Uniform Approximation for the $6j$-Symbol}

We now derive the uniform approximation for the $6j$-symbol using equation (\ref{ch5: eq_general_uniform}) and the canonical map $Z: (K_x, K_y, K_z) \rightarrow (J_x, J_y, J_z)$ constructed in section \ref{ch5: sec_symplectic_map}. The symbol Hamiltonian is $J_{23}$ from equation (\ref{ch5: eq_J_23_function}), and the normal form symbol is $J_n$, where 

\begin{equation}
{J}_n = \hat{\bf n} \cdot {\bf {J}} ={J}_z \cos \beta +{J}_y \sin \beta \, .
\label{ch5: eq_J_n_def}
\end{equation}

The coordinates $K_z$ and $J_z$ play the roles of the coordinates $x$ and $X$, respectively. From the construction of $Z$, we have 

\begin{eqnarray}
\label{ch5: eq_J_23_J_n_relation}
(J_{23}(K_z, \phi_{12}) - J_{23}) \circ Z^{-1} = A (J_n - m')  \, ,   \\
\label{ch5: eq_J_12_J_z_relation}
(J_{12}(K_z, \phi_{12}) - J_{12} ) \circ Z^{-1} = B (J_z - m)  \, , 
\end{eqnarray}
for some functions $A(J_z, \phi)$ and $B(J_z, \phi)$. The formula equation (\ref{ch5: eq_general_uniform}) gives 

\begin{equation}
\label{ch5: eq_uniform_formula_2}
\braket{j_{12} | j_{23}}  = \braket{J_z \, : \, m | \frac{1}{\sqrt{\hat{A} \hat{B}}}  | J_n \, : \, m'}  \, . 
\end{equation}
Because $\hat{A}$ and $\hat{B}$ are part of the amplitude, we only need to evaluate them to zeroth order. We assume that $A = f(J_z, J_n)$ and $B=g(J_z, J_n)$ can be expressed, respectively, as some functions $f$, $g$ of the variables $J_z$ and $J_n$, so to zeroth order, $\hat{A} = f(\hat{J}_z, \hat{J}_n)$ and $\hat{B} = g(\hat{J}_z, \hat{J}_n)$. Acting $\hat{J}_z$ on $\ket{J_z \, : \, m}$ on the left replaces the operator $\hat{J}_z$ by the eigenvalue $m$. Similarly, acting $\hat{J}_n$ on $\ket{J_n \, : \, m'}$ on the right replaces the operator $\hat{J}_n$ by $m'$. Here we have ignored the ordering of $\hat{J}_z$ and $\hat{J}_n$, since we are evaluating $\hat{A}$ and $\hat{B}$ to zeroth order. Thus equation (\ref{ch5: eq_uniform_formula_2}) becomes 

\begin{eqnarray}
\braket{j_{12} | j_{23}}  &=& \frac{1}{\sqrt{f(m, m') g(m, m')}}  \braket{J_z \, : \, m  | J_n \, : \, m'}  \nonumber  \\
&=& \frac{1}{\sqrt{A(m, \phi_0) B(m, \phi)}}  d^j_{mm'} (\beta)  \, , 
\label{ch5: eq_uniform_formula_3}
\end{eqnarray}
where the point $(J_z, \phi) = (m, \phi_0)$ is the intersection point of the level sets $J_z = m$ and $J_n = m'$, and $\phi = \phi_0$ is given by equation (51) of \cite{littlejohn2009}. 

From the definitions of $A$ in equation (\ref{ch5: eq_J_23_J_n_relation}), we can evaluate $A$ at the point $(m, \phi_0)$ as a limit $\phi \rightarrow \phi_0$ on the curve $K_z = m$,

\begin{eqnarray}
A(m, \phi_0)  &=& \lim_{\phi \rightarrow \phi_0} \frac{ J_{23} \circ Z^{-1} - J_{23} }{J_n(m, \phi) - m'}  \nonumber  \\
&=& \frac{ \partial J_{23}(K_z ( m, \phi), \phi_{12}(m, \phi))  / \partial \phi }{ \partial J_n(m, \phi) / \partial \phi}  \nonumber  \\
&=& \frac{ ( \partial J_{23}  / \partial \phi_{12} )  \, (\partial \phi_{12} / \partial  \phi)  }{ \partial J_n(m, \phi) / \partial \phi}  \nonumber \\
&=& \frac{  \{ J_{12}, J_{23}  \}  \, (\partial \phi_{12} /  \partial \phi)   }{ \{ J_z, J_n \} }  \, , 
\label{ch5: eq_6j_A_value}
\end{eqnarray}
where we have used L'Hoptal rule in the second equality, the chain rule and the fact $\partial K_z(m, \phi) / \partial \phi = 0$ in the third equality. In the fourth equality, we have used the fact that $(J_{12}, \phi_{12})$ and $(K_z, \phi)$ are canonical coordinates, so the derivatives can be replaced by Poisson brackets.

Similarly, we can evaluate $B$ from its definition in equation (\ref{ch5: eq_J_12_J_z_relation}) as a limit $J_z \rightarrow m$ on the curve $\phi = \phi_0$,

\begin{equation}
B(m, \phi_0) = \lim_{J_z \rightarrow m} \frac{ K_z \circ Z^{-1} - m  }{ J_z - m }  
=  \frac{ \partial K_z / \partial J_z}{\partial J_z / \partial J_z }  
= \frac{ \partial K_z}{\partial J_z}  \, . 
\label{ch5: eq_6j_B_value}
\end{equation}
It turns out the product of the two derivatives $\partial K_z / \partial J_z$ and $\partial \phi_{12} / \partial \phi$ is equal to the Jacobian of the canonical transformation at the intersection point $(m, \phi_0)$, which is equal to unity by the area preserving property of the map. To see this, note that $K_z(J_z = m, \phi) = m$ implies that $\partial K_z / \partial \phi = 0$ at $(m, \phi_0)$. So the Jacobian simplifies as follows: 

\begin{equation}
1 = \det
\left(
\begin{array}{cc}
\partial K_z / \partial J_z  &  \partial K_z / \partial \phi  \\
\partial \phi_{12} / \partial J_z  & \partial \phi_{12} / \partial \phi
\end{array}
\right)
= \, \left(   \frac{ \partial K_z}{\partial J_z}  \right)  \,   \left( \frac{\partial \phi_{12}}{ \partial  \phi } \right)   \, . 
\label{ch5: eq_unity_determinant}
\end{equation}
Thus, from equations (\ref{ch5: eq_6j_A_value}), (\ref{ch5: eq_6j_B_value}), and (\ref{ch5: eq_unity_determinant}), we find

\begin{equation}
\label{ch5: eq_AB_value}
A(m, \phi_0) \, B(m, \phi_0) = \frac{ \{ J_{12}, J_{23}  \} }{ \{ J_z, J_n \} }  \, . 
\end{equation}
Putting equation (\ref{ch5: eq_AB_value}) back into equation (\ref{ch5: eq_uniform_formula_3}), we have 

\begin{equation}
\label{ch5: eq_uniform_formula_4}
\braket{j_{12} | j_{23}} =  \left( \frac{ \{ J_z , J_n \} }{\{ J_{12} , \, J_{23} \}} \right)^{1/2} d^j_{mm'} (\beta)   \, . 
\end{equation}
Using the expression of $J_n$ from equation (\ref{ch5: eq_J_n_def}) , we find 
 
\begin{equation}
\label{ch5: eq_Jz_Jn_bracket}
| \{ J_z, J_n \} |
= \sin \beta \sqrt{J^2 - J_z^2} \sin \phi_0   
= \sin \beta J_y   \, , 
\end{equation}
where $J_y = \sqrt{J^2 - J_z^2}  \sin \phi_0$. Either through a direct calculation using the expression of $J_{23}$ from equation (\ref{ch5: eq_J_23_function}), or using equation (73) from \cite{littlejohn2010}, we have 

\begin{equation}
\label{ch5: eq_J12_J23_bracket}
| \{ J_{12}, J_{23} \} | = \frac{ \{ J_{12}^2, J_{23}^2 \} }{J_{12} \, J_{23} } = \frac{ 6V }{J_{12} \,  J_{23} }  \, , 
\end{equation}
where $V$ is the volume of the tetrahedron formed by the six edge lengths $J_1, J_2, J_3, J_4, J_{12}, J_{23}$. Finally, putting the Poisson brackets back into equation (\ref{ch5: eq_uniform_formula_4}) and using the definition of the $6j$-symbol from equation (\ref{ch5: eq_6j_symbol_def}), we arrive at the uniform approximation of the $6j$-symbol in terms of the $d$-matrices

\begin{eqnarray}
  \label{6j_result}
    \left\{ 
   \begin{array}{ccc} 
     j_1 & j_2 & j_{12} \\
     j_3 & j_4 & j_{23} 
   \end{array}
   \right\} 
   &=& (-1)^{\gamma} \, \frac{1}{\sqrt{4 J_{12} \, J_{23}}} \, \left( \frac{ | \{ J_z , J_n \} | }{ | \{ J_{12} , \, J_{23} \} | } \right)^{1/2} d^j_{mm'} (\beta)  \nonumber  \\
   &=& (-1)^{\gamma} \, \left(\frac{ | \sin \beta \, J_y | }{ | 24V | } \right)^{1/2} \; d^j_{mm'} (\beta) \, . 
   \label{ch5: eq_uniform_formula_5}
\end{eqnarray}
where we have put back an arbitrary phase $(-1)^{\gamma}$ that we have been ignoring up to now. This extra phase is given in equation (69) and equation (75) of \cite{littlejohn2009}. The accuracy of the uniform approximation in equation (\ref{ch5: eq_uniform_formula_5}) is excellent. It is superior compared to the Ponzano Regge formula even in the classically allowed regions. See \cite{littlejohn2009} for more details on the numerical performance of this uniform approximation.

\section{Conclusion}

In this paper, we found that the symbol of the $\hat{J}_{23}$ matrix operator is the edge length $J_{23}$ in a tetrahedron as a function of the opposite edge length $J_{12}$ and dihedral angle $\phi_{12}$. This coincidence, though not surprising, has allowed us to use the Schl\"{a}fli identity to derive the Ponzano-Regge phase for the area of the lune in figure \ref{ch5: fig_canonical_map_1}. 

There is a similar matrix operator $\hat{J}_{23}$ for the $q$-deformed $6j$-symbol, and a similar Schl\"{a}fli identity for spherical tetrahedra. If we can show that the symbol of $\hat{J}_{23}$ in the $q$-deformed $\ket{j_{12}}$ basis is the $J_{23}$ function in a spherical tetrahedron, we can use the method in this paper to give a new derivation for the asymptotic formula of the $q$-deformed $6j$-symbol. We can also derive a new uniform approximation in terms of the rotation matrices. We shall pursue this avenue of research in a future project. 

Another application of the method developed in this paper is in generalizing the Bohr-Sommerfeld quantization conditions from Schr\"{o}dinger equations to matrix operators. Given that the level sets of the symbol of the eigenvalue equations are mapped to the level sets of the $d$-matrices, which contain quantized areas on the sphere, we can derive quantization conditions by quantizing the area enclosed by the level set of the symbol of the original matrix operator. A recent example of quantization conditions on spherical phase space is the Bohr-Sommerfeld quantization condition \cite{haggard2011} for the volume operator.

\appendix

\section{\label{ch5: section_unitary_operator}Construction of the Unitary Operator}

Although we do not actually need to construct the unitary operator $\hat{U}$ explicitly in our calculation, for completeness, we now give details on the construction of $\hat{U}$ from the canonical map $Z$ in this appendix. This discussion will also make it clear that $Z^{-1}$, not $Z$, should be used in equation (\ref{ch5: eq_normal_form_operator}).

Given a Hamiltonian $\hat{H}$ and a desired normal form $K_0$, we assume there exist a canonical map $Z$ of the phase space that transforms $H_0$ to $K_0$, in other words, $ K_0 = H_0 \circ Z^{-1} $. We imbed $Z$ into a smooth one parameter family of canonical maps $Z_\epsilon\,$, $\epsilon\in[0,1]$, with the boundary condition $Z_0 = I$ and $Z_1 = Z$.  We set the symbol of the generators of the unitary transformation $G_\epsilon$ to be the solution of the differential equation

\begin{equation}
  \label{dZde}
  \frac{dZ^\mu_\epsilon}{d\epsilon} = \{Z^\mu_\epsilon, G_\epsilon\} \, . 
\end{equation}
Once we have $G_\epsilon$, we use the inverse symbol map to find $\hat{G}_\epsilon$, and construct $\hat{U}_\epsilon$ as the solution of the operator equation

\begin{equation}
  \label{dUde}
  \frac{d\hat{U}_\epsilon}{d\epsilon} = - \frac{i}{\hbar} \, \hat{U}_\epsilon \hat{G}_\epsilon \, ,
\end{equation}
with boundary condition $\hat{U}_0=\hat{I}$. We finally set $\hat{U} = \hat{U}_1$. 

To show that $K_0 = H_0 \circ Z^{-1}$ is indeed the principal symbol of $\hat{K} = \hat{U} \hat{H} \hat{U}^\dagger$. We will trace the evolution of $\epsilon$ backwards from $\hat{K}$ to
$\hat{H}$. Suppose $\hat{K}$ is known. Denote the symbol of $\hat{K}$ up to order $\hbar$ by $K_0'$. We want to show that $K_0' = K_0$. Define $\hat{H}_\epsilon= \hat{U}_\epsilon^\dagger \hat{K} \hat{U}_\epsilon$, so that $\hat{H}_0 = \hat{K}$, $\hat{H}_1 = \hat{H}$. Differentiate $\hat{H}_\epsilon$ and use equation (\ref{dUde}) to get

\begin{equation}
  \frac{d\hat{H}_\epsilon}{d\epsilon} = \frac{i}{\hbar} [\hat{G}_\epsilon , \hat{H}_\epsilon] \, .
\end{equation}
Transcribing to symbols using equation (\ref{ch5: eq_commutator_bracket}), keeping terms up to first order in $\hbar$, we find

\begin{equation}
  \frac{dH_\epsilon}{d\epsilon} = - \{G_\epsilon, H_\epsilon\}  \, .
\end{equation}
Then $K_0' \circ Z_\epsilon$ is the solution to the above equation, as a result of the following calculation:

\begin{eqnarray}
\frac{d}{d\epsilon} ( K_0' \circ Z_\epsilon )  
&=& \frac{dK_0'}{d\epsilon} \circ Z_\epsilon + (K_{0,\mu}' \circ Z_\epsilon) \, \frac{dZ^\mu_\epsilon}{d\epsilon}  \nonumber \\ 
&=& (K_{0,\mu}' \circ Z_\epsilon) \{Z^\mu_\epsilon, G_\epsilon\}  \nonumber \\
&=& \{ K_0' \circ Z_\epsilon , G_\epsilon \}  \, , 
\end{eqnarray}
where we have used equation (\ref{dZde}) in the second equality, and the chain
rule in the third equality. Thus, we find $H_\epsilon = K_0' \circ Z_\epsilon$. 

The boundary condition $H_0 = K_0' \circ Z_\epsilon$ at $\epsilon = 1$ then requires $K_0' = H_0 \circ Z^{-1}$, which is equal to the normal form $K_0$ by our choice of the canonical map $Z$. This shows the symbol of $\hat{K} =\hat{U} \hat{H} \hat{U}^\dagger $ up to order $\hbar$ is in the required normal form $K_0$.  This completes the demonstration of the relationship between unitary transformations of the Hamiltonian operator and the canonical transformations of classical Hamiltonian functions.


\section*{References}

\bibliographystyle{plain}

\begin{thebibliography}{10}

\bibitem{littlejohn2010}
V.~Aquilanti, H.~M. Haggard, A.~Hedeman, N.~Jeevanjee, R.~G. Littlejohn, and
  L.~Yu.
\newblock Semiclassical mechanics of the {Wigner} $6j$-symbol.
\newblock {\em e-Print}, arXiv: math-ph/1009.2811, 2010.

\bibitem{berry1972}
M.~V. Berry and K.~E. Mount.
\newblock Semiclassical approximations in wave mechanics.
\newblock {\em Rep. Prog. Phys.}, 35:315, 1972.

\bibitem{haggard2011}
E.~Bianchi and H.~M. Haggard.
\newblock Discreteness of the volume of space from {Bohr-Sommerfeld}
  quantization.
\newblock {\em Phys. Rev. Lett.}, 107:011301, 2011.

\bibitem{cargo2005b}
M.~Cargo, A.~Gracia-Saz, and R.~G. Littlejohn.
\newblock Multidimensional quantum normal forms, {Moyal} star product, and
  torus quantization.
\newblock {\em e-Print}, arXiv: math-ph/0507032, 2005.

\bibitem{cargo2005a}
M.~Cargo, A.~Gracia-Saz, R.~G. Littlejohn, M.~W. Reinsch, and P.~de~M.~Rios.
\newblock Quantum normal forms, {Moyal} star product and {Bohr-Sommerfeld}
  approximation.
\newblock {\em J. Phys. A}, 38:1977, 2005.

\bibitem{cargo2002}
M.~Cargo and R.~G. Littlejohn.
\newblock Phase space deformation and basis set optimization.
\newblock {\em Phys. Rev. E}, 65:026703, 2002.

\bibitem{edmonds1960}
A.~R. Edmonds.
\newblock {\em Angular Momentum in Quantum Mechanics}.
\newblock Princeton University Press, Princeton, 1960.

\bibitem{freidel2002}
L.~Freidel and K.~Krasnov.
\newblock The fuzzy sphere star-product and spin networks.
\newblock {\em J. Math. Phys.}, 43:1737, 2002.

\bibitem{varilly1988}
J.~M. Gracia-Bondia and J.~C. Varilly.
\newblock The {Moyal} representation for spin.
\newblock {\em Ann. Phys.}, 190:107, 1989.

\bibitem{littlejohn2009}
R.~G. Littlejohn and L.~Yu.
\newblock Uniform semiclassical approximation for the {Wigner} $6j$ symbol in
  terms of rotation matrices.
\newblock {\em J. Phys. Chem. A}, 113:14904, 2009.

\bibitem{milnor1994}
J.~Milnor.
\newblock The {Schl\"{a}fli} differential equality.
\newblock In {\em Collected Papers, Vol. 1}. Publish or Perish, Houston, 1994.

\bibitem{schulten1975a}
K.~Schulten and R.~G. Gordon.
\newblock Exact recursive evaluation of $3j$- and $6j$-coefficients for
  quantum-mechanical coupling of angular momenta.
\newblock {\em J. Math. Phys.}, 16:1961, 1975.

\bibitem{stratonovich1956}
R.~L. Stratonovich.
\newblock On distributions in representation space.
\newblock {\em Sov. Phys. JETP}, 4:891, 1957.

\end{thebibliography}

\end{document}